\documentclass[prl,twocolumn,showpacs]{revtex4}
\usepackage[dvips]{graphicx}
\usepackage{latexsym}
\usepackage{bm}
\begin{document}
\newcommand{\fig}[2]{\includegraphics[width=#1]{#2}}
\newcommand{\pprl}{Phys. Rev. Lett. \ }
\newcommand{\pprb}{Phys. Rev. {B}}
\newcommand{\be}{\begin{equation}}
\newcommand{\ee}{\end{equation}}
\newcommand{\bea}{\begin{eqnarray}}
\newcommand{\eea}{\end{eqnarray}}
\newcommand{\nn}{\nonumber}
\newcommand{\la}{\langle}
\newcommand{\ra}{\rangle}
\newcommand{\dg}{\dagger}
\newcommand{\upa}{\uparrow}
\newcommand{\dna}{\downarrow}

\title{Itinerant and localized magnetism on the triangular lattice:
\\ sodium rich phases of Na$_x$CoO$_2$}

\author{Meng Gao, Sen Zhou, and Ziqiang Wang}
\affiliation{Department of Physics, Boston College, Chestnut Hill,
MA 02467}

\date{\today}

\begin{abstract}
We study the interplay between correlation, itinerant ferromagnetism
and local moment formation on the electron doped triangular lattice of
sodium cobaltates Na$_x$CoO$_2$. We find that strong
correlation renormalizes the Stoner criterion and stabilizes the
paramagnetic state for $x<x_c\simeq0.67$. For $x>x_c$, ferromagnetic
(FM) order emerges. The enhanced Na dopant potential fluctuations play
a crucial role in the sodium rich phases and lead to
an inhomogeneous FM state, exhibiting nonmagnetic
Co$^{3+}$ patches, antiferromagnetic (AF) correlated regions, and FM
clusters with AF domains. Hole doping the band insulator at x=1 leads to
the formation of local moments near the Na vacancies and AF
correlated magnetic clusters. We explain
recent observations by neutron, $\mu$SR, and NMR experiments
on the evolution of the magnetic properties in the sodium rich phases.

\typeout{polish abstract}
\end{abstract}


\maketitle

Sodium cobaltates, Na$_x$CoO$_2$, are electron doped transition metal oxides
with a layered, hexagonal lattice structure. The Co$^{4+}$ has a 3d$^9$
configuration with 5 d-electrons occupying the three lower $t_{2g}$ orbitals.
At a Na density $x$, the average Co valence is Co$^{(4-x)}$, evolving from
an open shell Co$^{4+}$ with low-spin state $S=1/2$ at $x=0$ to a closed shell
Co$^{3+}$ with a low-spin state $S=0$ at $x=1$. The cobaltates provide an
almost ideal low-spin lattice fermion system for studying the physics of
correlation and geometrical frustration in the charge, spin, and orbital
sectors.

Since the discovery of unconventional superconductivity near $x=0.3$
upon water-intercalation in this material \cite{takada03}, a broad
spectrum of experiments have been performed, yielding a rich and
complex phase diagram with many unexpected and novel properties.
Magnetism plays an essential role in the sodium rich region with
$x>0.5$. The metallic transport coexists with a Curie-Weiss magnetic
susceptibility leading to a novel ``Curie-Weiss metal''
\cite{foo04}. Neutron scattering and NMR experiments find strong
in-plane ferromagnetic (FM) fluctuations at high temperatures and
magnetically ordered states emerge for $x>0.75$ \cite{boothroyd}. At
$x=0.82$, neutron scattering determined that in-plane FM order and
inter-layer antiferromagnetic (AF) order develops below a 3D Neel
temperature $T_N=20$K \cite{bayrakci}. There is increasing
experimental evidence for unexpected strong correlation effects as
the cobaltates approach the band insulting limit at $x=1$. Recent
$\mu$SR experiments discovered the emergence of AF correlated
magnetic clusters and localized magnetic moments for $x>0.96$
\cite{musr}. Interestingly, the high thermoelectric power of the
sodium cobaltate is found in the sodium rich region which is likely
to have a related magnetic origin \cite{ong06}. The nature of the
magnetism in the sodium rich region has been the focus of several
theoretical work \cite{powell,shastry,korshunov,khaliullin}.

In this paper, we study the magnetic
properties, both itinerant and localized, of strongly correlated electrons
on the triangular lattice, and present a theoretical description of the
novel magnetism observed by experiments in the sodium rich phases.
Strong correlation plays an important role
in the electronic structure of the cobaltates.
It has been shown recently \cite{zhou05,liebsch,marietti,bourgeois} that
correlation-induced corrections to the crystal field splitting and
bandwidths of the LDA calculations \cite{singh} can produce a single
renormalized band of mostly $a_{1g}$ character at the Fermi level as
observed by angle resolved photoemission (ARPES)
\cite{hasan,hbyang04,hbyang05,hasan06}.
Here we show that the description of the magnetism
in the cobaltates, too, requires a proper account of the large
Coulomb repulsion
$U$ at the cobaltate sites. Indeed, in the LSDA+U theory \cite{zhang},
which accounts for the correlation effects
in the Hartree-Fock approach, the LDA paramagnetic (PM) state
is always unstable and the ground state is a fully
polarized ferromagnet at all doping for $U$ as small as $2$eV.
This disagrees with experiments that the in-plane FM order does not
emerge until $x>0.75$ and that a single spin-degenerate Fermi surface
of the Luttinger area is observed by ARPES for $x<0.72$ \cite{hbyang05}.
We show that the weak-coupling Stoner instability is unphysical when
correlation is strong. Using a strong-coupling Gutzwiller projection approach,
we show that the Stoner criterion is strongly renormalized and the spin
dependent self-energy scales with the average kinetic energy instead of $U$.
The PM phase is in fact stable against itinerant FM
below a critical electron doping $x_c\simeq0.67$ above which
in-plane FM order emerges.

In the sodium-rich region, we show that the enhanced electrostatic
potential fluctuations due to the disordered Na dopants lead to
the coexistence of localized and itinerant electronic states with
inhomogeneous FM order, exhibiting nonmagnetic Co$^{3+}$ patches,
AF correlated local regions, and FM clusters with AF domain walls.
For very high Na doping, the dilute Na vacancies enhance the strong
correlation effects by increasing the localization tendency of the carriers
\cite{markotliar}. We consider the case of a few holes (Na vacancies)
doped into the band insulator at $x=1$, and find that the hidden correlation
effects are brought out upon the slightest amount of doping.
Specifically, a single hole/Na vacancy
induces an $S=1/2$ local moment. We address the interactions between
the local moments and the evolution into magnetic clusters and eventually
to the macroscopic FM ordered state with decreasing Na doping and provide
a theoretical description of neutron \cite{bayrakci}, $\mu$SR \cite{musr}
and NMR experiments \cite{mukhamedshin07}.

We start with the Hubbard model for the relevant low-energy
quasiparticle band of approximate $a_{1g}$ character, \be
H=\sum_{i,j,\sigma} t_{ij} c_{i\sigma}^{\dagger}c_{j\sigma}
+\sum_{i}U\hat{n}_{i\uparrow}\hat{n}_{i\downarrow}, \label{h} \ee
where $c_{i\sigma}^\dagger$ creates a {\em hole} and $U$ ($\sim5$eV
for the cobaltates \cite{hasan}) is the on-site Coulomb repulsion.
To model the $a_{1g}$ band, we consider up to 3rd nearest-neighbor
hopping ($t_1,t_2,t_3$)=($-0.202,0.035,0.029$)eV \cite{zhouwang}.
The hole density $n_i=n_{i\uparrow}+n_{i\downarrow}=1-x_i$ with
$x_i$ the electron doping, is fixed by the chemical potential $\mu$.
Due to the small direct Co-Co overlap and the large $U$ and the
$90^\circ$ O-Co-O bond angle, the AF superexchange $J$ in the
cobaltates is small \cite{motrunichlee}, consistent with the value
$J\sim5$meV determined by inelastic neutron scattering
\cite{bayrakci}. We thus focus on the in-plane magnetism of the
kinetic origin. To study the 3D magnetic ordering transition at
finite temperature, a small interlayer exchange coupling is needed.

In the weak-coupling HF theory, itinerant FM is due to the Stoner
instability, i.e. the divergence of the uniform susceptibility
$\chi=\chi_0/(1-UN_F/2)$, where $\chi_0$ is the free-electron value
and $N_F$ is the DOS at the Fermi level. The large DOS of the
cobaltates would lead to FM order for all $x$ for $U$ greater than a
value less than 2eV \cite{zhang}, clearly inconsistent with
experiments. The failure of the HF or LSDA+U theory lies in the
spin-dependent self-energy correction that scales with $U$. This is
unphysical for $U$ larger than the bandwidth. For example, for large
$U$, the system can simply avoid paying the energy penalty for
double occupation in Eq.~(\ref{h}) by reverting to a fully
spin-polarized state that involves only the kinetic energy. This is
the physics behind the Nagaoka theorem: on square lattices, the
ground state of the infinite-U Hubbard model doped with a single
hole is a fully polarized FM. At finite hole density, the Nagaoka
state is lower in energy than the Gutzwiller projected PM state at
low doping \cite{kr}. Interestingly, on the triangular lattice the
kinetic energy is frustrated in the sense that hopping around an
elemental triangle picks up a negative sign, the Nagaoka state is
not the ground state for a single hole \cite{shastryonehole}. To
study FM at finite electron doping, we consider the large-U limit
captured by the projection of double occupation. To make analytical
progress, we treat the latter by Gutzwiller approximation (GA) in
the grand canonical ensemble where the projected wave function can
be written as \be \vert\Psi\rangle=\prod_i y_{i\uparrow}^{\hat
n_{i\uparrow}} y_{i\downarrow}^{\hat n_{i\downarrow}}(1-{\hat
n_{i\uparrow}} {\hat n_{i\downarrow}})\vert\Psi_0\rangle. \label{wf}
\ee Here $\Psi_0$ is an unprojected Slater determinant state and
$y_{i\sigma}$ is a spin-dependent local fugacity that maintains the
equilibrium and the local densities upon projection. The GA is then
equivalent to minimizing the energy of the renormalized Hamiltonian,
\be H_{GA}= \sum_{i,j,\sigma} g_{ij}^\sigma
t_{ij}c_{i\sigma}^{\dagger}c_{j\sigma} +\sum_{i,\sigma}
\varepsilon_{i\sigma}(c_{i\sigma}^{\dagger}c_{i\sigma}-n_{i\sigma}),
\label{hga} \ee where the Gutzwiller renormalization factor, \be
g_{ij}^\sigma={\langle\Psi\vert c_{i\sigma}^\dagger
c_{j\sigma}\vert\Psi\rangle\over\langle\Psi_0\vert
c_{i\sigma}^\dagger c_{j\sigma}\vert\Psi_0\rangle}\simeq\sqrt{x_i
x_j\over(1-n_{i\sigma}) (1-n_{j\sigma})}, \label{gij} \ee and
$\varepsilon_{i\sigma}$ is determined by $\partial\langle
H_{GA}\rangle/\partial n_{i\sigma}=0$, \be
\varepsilon_{i\sigma}={1\over2(1-n_{i\sigma})} \sum_j g_{ij}^\sigma
t_{ij} \langle c_{i\sigma}^\dagger c_{j\sigma} +h.c.\rangle -
\varepsilon_{i0}. \label{epsilon-sigma} \ee Physically,
$\varepsilon_{i\sigma}$ is the local kinetic energy per doped
spin-$\sigma$ electron measured relative to the average over both
spins,
$\varepsilon_{i0}=-\sum_\sigma(1-n_{i\sigma})\varepsilon_{i\sigma}$.
\begin{figure}
\begin{center}
\fig{2.8in}{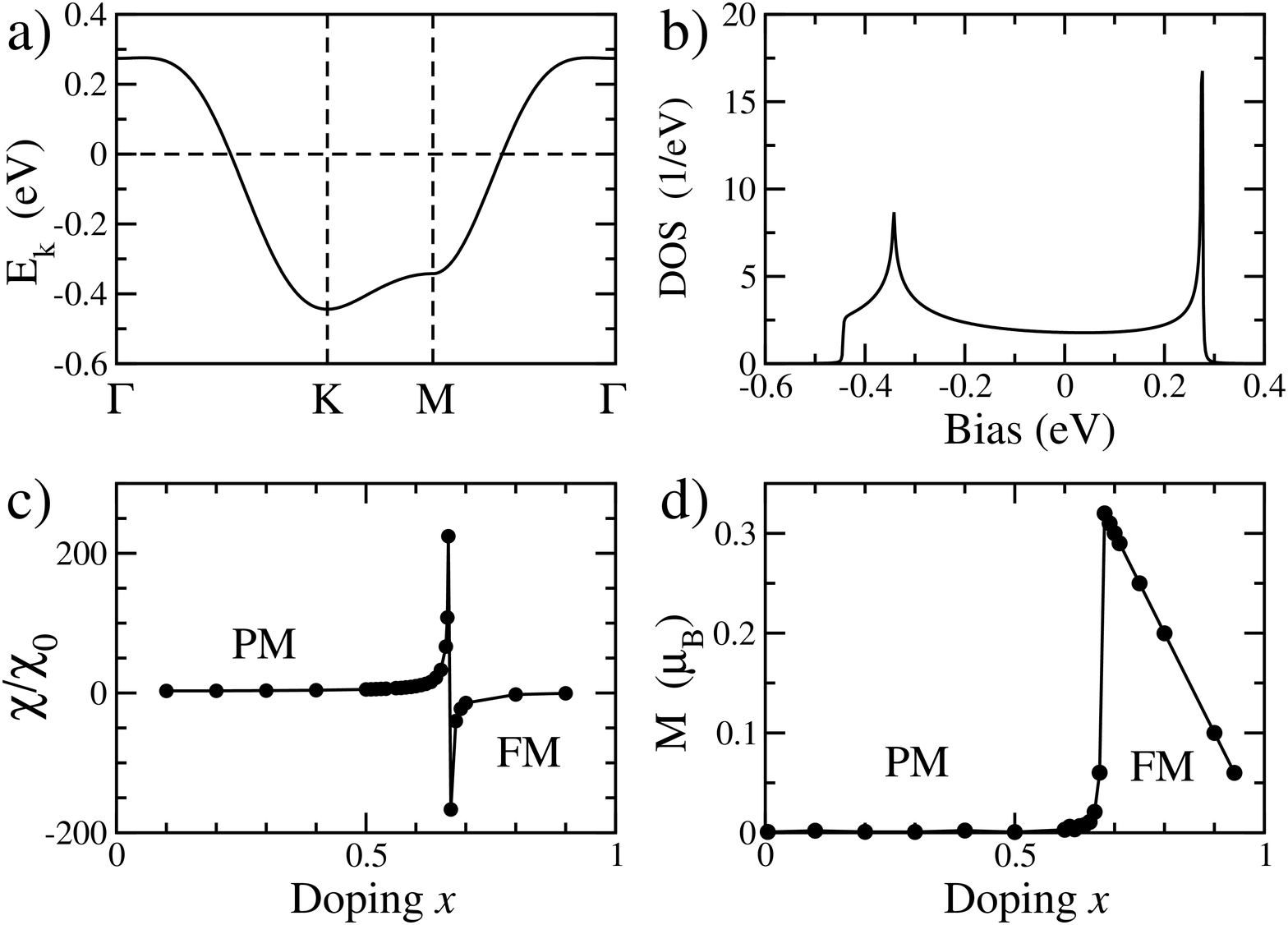}
\vskip-2.4mm
\caption{
Renormalized $a_{1g}$ quasiparticle dispersion (a) and
the DOS (b) in the PM state at x=0.3. (c) The uniform susceptibility
showing the renormalized Stoner instability at $x_c\simeq0.67$. (d)
The magnetization phase diagram.
}
\label{fig1}
\end{center}
\vskip-9mm
\end{figure}

In the uniform phase, $n_{i\sigma}=n_\sigma$,
$g_{ij}^\sigma=g^\sigma$, $\varepsilon_{i\sigma}=\varepsilon_\sigma$,
\be
\varepsilon_\sigma={1\over(1-n_\sigma)}\sum_k
E_{k\sigma} f(E_{k\sigma}+\varepsilon_\sigma-\mu)-\varepsilon_0,
\label{epsilon}
\ee
where
$f$ is the Fermi function and $E_{k\sigma}=g^\sigma[
2t_1(\cos k_y+2\cos\frac{\sqrt{3}k_x}{2}\cos\frac{k_y}{2})
+2t_2(\cos\sqrt{3}k_x+2\cos\frac{\sqrt{3}k_x}{2}\cos\frac{3k_y}{2})
+2t_3(\cos2k_y+2\cos\sqrt{3}k_x\cos k_y)]$ is
the renormalized quasiparticle dispersion.
The quasiparticle band and the DOS
in the PM phase are shown in Fig.~1a at $x=0.3$.
The bandwidth reduction is due to the Gutzwiller factor
$g^\sigma <1$. The peak in the DOS near the band top is a property
of the cobaltate $a_{1g}$ band captured with up to third n.n. hopping.

To drive a renormalized Stoner theory for the instability of the PM
phase against uniform FM order, we calculate the uniform magnetic
susceptibility $\chi=\partial M/\partial h$, where
$M=(n_\uparrow-n_\downarrow)\mu_B$ is the magnetization due to an
infinitesimal external magnetic field $h$. Including the leading
$h$-dependence of $\varepsilon_\sigma$ in Eq.~(\ref{epsilon}) we
obtain \be \chi={\chi_0\over 1-KN_F},\quad K=-{2\over
1+x}\left(E_{k_F} +\varepsilon+\varepsilon_0\right), \label{chi} \ee
where $N_F$ is the renormalized DOS and $K$ measures the energy per
electron at the Fermi level. The latter plays the role of the
effective interaction strength, replacing the Hubbard-U in the
Stoner susceptibility. Fig.~1b shows the doping dependence of $\chi$
which diverges at $x_c\simeq0.67$, corresponding to the renormalized
Stoner criterion $KN_F=1$, beyond which FM order develops at $T=0$.
The self-consistently determined magnetization shown in Fig.~1c
indicates a sharp transition into the fully polarized FM,
half-metallic state.
The presence of the DOS peak near the band top is
important for the emergence of the in-plane FM order. For example,
with only nearest neighbor hopping, the FM order is
completely suppressed by strong correlation \cite{powell}.
We obtained the same results from the
three-band Hubbard model of the $t_{2g}$ complex.
The finding of the fully polarized FM state at large $x$ is
consistent with the large FM moment of about
$0.13\mu_B$ per Co site at $x=0.82$ observed by neutron
scattering \cite{bayrakci} where $0.18\mu_B$ corresponds to a fully
polarized FM state.
\begin{figure}
\begin{center}
\fig{3.4in}{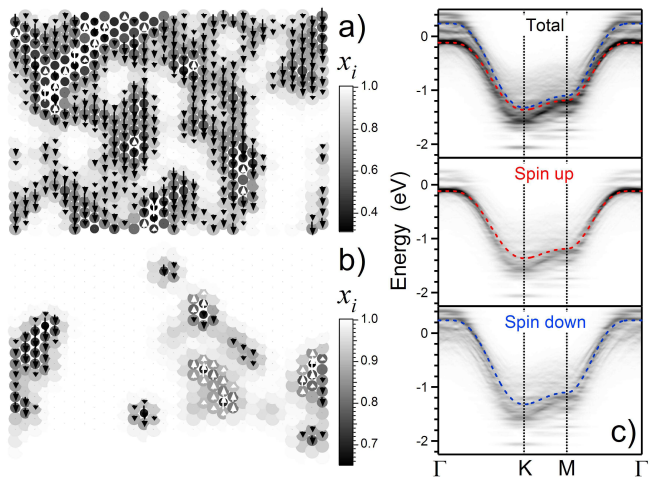} \vskip-2.4mm \caption{ (Color online) The spin (black and
white arrows) and charge (local doping $x_i$) distribution in a
typical Na(1) and Na(2) realization at $x=0.8$ (a) and $x=0.97$. (c)
The one-particle spectral intensity (top) and its spin resolved
components at $x=0.8$. } \label{fig2}
\end{center}
\vskip-9mm
\end{figure}

Now we turn to the spatially unrestricted solutions of
Eqs.~(\ref{hga}-\ref{epsilon}) and study the effects of the Na
dopants. It is known that Na orders at $x=0.5$ into
$\sqrt{3}\times2$ superstructure \cite{zandbergen}, which has been
shown \cite{zhouwang} to play an important role in alleviating
geometrical frustration for the emergence of the $2\times2$ AF
ordered state observed by neutron scattering \cite{gasparovic}. For
$x\neq0.5$, Na ions are disordered, and occupy randomly the
preferred Na(1) and Na(2) sites directly above/below the Co or the
middle of the Co triangle respectively, with a ratio of $~1:7$ at
$x=0.8$ \cite{zandbergen}. The electrostatic potential is described
by adding to the Hamiltonian (\ref{h}), \be H_V=V\sum_{i>j}{\hat n_i
\hat n_j\over |\vec r_i-\vec r_j\vert}
+V_d\sum_i\sum_{\ell=1}^{N_{\rm N_a}}{\hat n_i\over\sqrt {\vert\vec
r_\ell-\vec r_i\vert^2 +d_z^2}}, \label{v} \ee where $V_d$ is the
dopant potential strength, $d_z\simeq a$ the setback distance of Na
to the Co plane, and $V$ the long-range Coulomb interaction that
must be included to account for carrier screening. Figs.~2a and 2b
show the typical charge and spin distribution of the inhomogeneous
FM state at $x=0.8$ and $x=0.97$ on triangular lattices of
$32\times20$ sites, with $V=0.2$eV and $V_d=0.6$eV. Note that at
such strength of $(V,V_d)$, the ordered Na at $x=0.5$ only induces a
weak charge modulation \cite{zhouwang}. In contrast, the random
distribution of Na at $x=0.8$ leads to large fluctuations in the
local electrostatic potential, causing the localization of the
electrons and the formation of the nonmagnetic Co$^{3+}$ clusters
where $x_i\simeq1$. This result is in line with NMR: the presence of
Co$^{3+}$ for $x>0.65$ but not at $x=0.5$
\cite{bobroff06,mukhamedshin07}. The one-particle spectral intensity
in Fig.~2c clearly demonstrates the coexistence of localized and
itinerant band-like states. Interestingly, AF correlated regions
emerge at $x=0.8$ in locally underdoped regions in Fig.~2a. In these
regions, $x$ is much smaller and the kinetic AF correlation imbedded
in the Gutzwiller factor in Eq.~(3) prevails due to the alleviated
AF frustration by charge inhomogeneity, as proposed for the ``0.5
phase'' \cite{zhouwang}. As a result, the average magnetic moment in
Fig.~2a is reduced from the fully polarized value to about
$0.13\mu_B$ per Co site, in qualitative agreement with the finding
of neutron scattering \cite{bayrakci}. For stronger potential
fluctuations, FM clusters with AF domain walls emerge, as seen more
prominently at $x=0.97$ in Fig.~2b. These glassy behaviors arise
because the localized states formed out of the majority spin band
are pushed above $E_F$ and occupied by holes, as shown in the
spin-resolved spectral intensity in Figs.~2c.

To further study the localized magnetism, we consider the limit
$x=1$ and ask what happens when a few holes are doped into the band
insulating NaCoO$_2$ by the Na vacancy.
The hidden correlation effects are brought out by the slightest
amount of doping. Fig.~3a displays the case of a single hole added
by a Na(2) vacancy. Instead of adding the hole into the minority of
the spin-polarized bands, localized states are created and pinned
near the Fermi level (Fig.~3a) to accommodate the doped hole,
leading to the formation of the spin-1/2 local moment distributed
near the Na vacancy as shown in Fig.~3b for Na(1) and Na(2)
respectively. The localized states induced by the Na electrostatic
potential \cite{markotliar} are spin-split by
$\Delta_i=\varepsilon_{i\uparrow}-\varepsilon_{i\downarrow}
=(1/2x_i)\sum_{j}g_{ij}^\uparrow \langle t_{ij}c_{i\uparrow}^\dagger
c_{j\uparrow}+h.c. \rangle$. The latter has a localized profile
whose amplitude $E_b=\Delta_{\rm max}-\Delta_{\rm min}$ is used as a
measure of the binding energy of the local moment. In Fig.~3c, $E_b$
is plotted as a function of the {\em bare} Na potential $V_d$. The
local moment develops ($E_b>0$) for $V_d$ as small as $0.2$eV.
Because the Na(1) vacancy is directly above a Co site, $E_b$ is
enhanced, making it easier for the local moment to form near Na(1)
vacancies. For even smaller $V_d$, it becomes difficult for our
finite size numerical calculations to discern the localized states
and the values of $E_b$, leaving open the possibility of
self-trapped, spontaneous local moment formation without the Na
potential.
\begin{figure}
\begin{center}
\fig{3.2in}{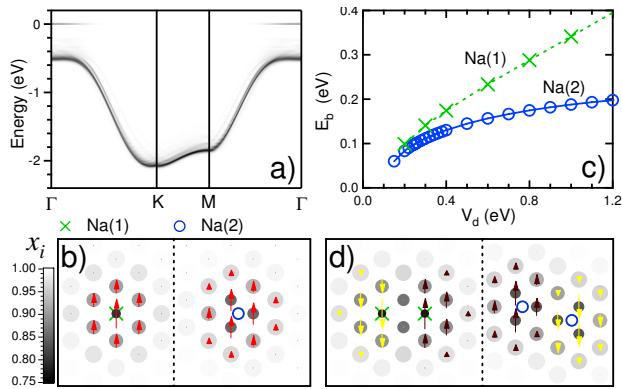}
\vskip-2.4mm
\caption{ (Color online)
(a)Spectral intensity of one-hole doped by
a Na(2) vacancy at $V_d=1.0$eV.
(b) Spin and charge distribution of the
S=1/2 local moment near a Na(1) and a Na(2)
vacancy. (c) The local moment ``binding energy'' $E_b$
as a function of $V_d$ induced by a Na(1) or Na(2) vacancy
at $V_d=0.6$eV.
(d) Spin/charge distribution around two Na(1) and
two Na(2) vacancies showing AF correlations between
the local moments.
}
\label{fig3}
\end{center}
\vskip-9mm
\end{figure}

When the Na vacancies are isolated, the spin-1/2 local moments
behave as free moments, contributing to significant spin entropy in
the Na-rich part of the phase diagram. As the vacancy density
increases, i.e. as the average $x$ reduces, the local moments begin
to overlap and their interactions become important. We find that
nearby local moments have AF correlations. As shown in Fig.~3d,
since the AF frustration is alleviated by the associated charge
inhomogeneity, two Na(2) or Na(1) vacancies induce two AF correlated
local moments with zero net magnetization. As the local density of
Na vacancy increases, FM clusters develop which eventually evolve
into the macroscopic FM state. Our findings provide a complimentary
description of the evolution from local magnetic clusters to
macroscopic FM state observed by recent $\mu$SR experiments
\cite{musr} and interpreted in terms of a change in the Co$^{3+}$
spin state that involves the higher Co-3d $e_g$ orbitals
\cite{bernhard04,khaliullin}.

We conclude with a discussion of an outstanding puzzle in the ARPES
experiments. A large hole-like FS is observed at $x=0.8$ with a volume
much larger than what is expected by the Luttinger counting in a
PM state \cite{qian06,ding07}. A natural explanation is that this
corresponds to the minority spin band of the itinerant half-metallic
state with in-plane FM order. Such an interpretation, although in line
with that of the neutron scattering experiments \cite{bayrakci}, implies a
filled majority band below the Fermi level which has yet to be detected
by ARPES. Our study suggests another
possible scenario where the in-plane FM order associated with the 3D A-type
AF order does not develop on the 2D surface at finite temperatures.
The larger than expected FS comes instead from the loss of the
doped electrons due to Na-induced localization and are ``taken out'' of the
$t_{2g}$ bands. For example, a substantial fraction of the doped electrons
on the
surface can be localized to form the nonmagnetic Co$^{3+}$ similar to Fig.~2a.
This picture is appealing given the recent finding by NMR of
valence disproportionation associated with significant Co$^{3+}$
formation for $0.65 < x < 0.75$
\cite{mukhamedshin07}. Interestingly, the same picture
within the present theory suggests that in the extremely sodium rich phases
$x>0.96$, it is the localization of a fraction of holes near the Na vacancies
that gives rise to the local magnetic clusters
observed by $\mu$SR experiments \cite{musr}. Further studies are
clearly needed to better understand the coexistence of itinerant
and localized magnetism in the sodium rich cobaltates.

We thank H. Ding and P. A. Lee for many useful discussions.
This work is supported by DOE grant DE-FG02-99ER45747.

\end{document}